\documentclass[pdftex,twocolumn,epjc3]{svjour3}          

\RequirePackage[T1]{fontenc}

\smartqed  

\usepackage[T1]{fontenc}
\usepackage{lmodern}
\usepackage{amsmath}
\usepackage{amsfonts}
\usepackage{amssymb}
\usepackage{mathrsfs}
\usepackage{physics}
\usepackage{graphicx}
\usepackage{caption}
\usepackage{subcaption}
\usepackage{enumitem}
\usepackage{framed,float}
\usepackage{booktabs}
\usepackage{array}
\usepackage{soul}
\usepackage{multirow}
\usepackage{comment}
\usepackage{braket}
\usepackage{todonotes}
\usepackage{youngtab}
\usepackage{tikz}
\usepackage{color}
\usepackage{extarrows}
\usetikzlibrary{calc,arrows,decorations.markings}
\usepackage{CJKutf8}
\usepackage{hyperref}

\renewcommand{\makeheadbox}{}

\def \be  {\begin{equation}}
\def \ee  {\end{equation}}
\def \ba {\begin{equation}\begin{aligned}}
\def \ea {\end{aligned}\end{equation}}
\def \bea  {\begin{eqnarray}}
\def \eea  {\end{eqnarray}}

\journalname{Eur. Phys. J. C}

\begin{document}


\title{\boldmath Predictions on observing hot holographic quark star with gravitational waves}

\author{Le-Feng Chen\thanksref{e1,addr1}
        \and
        Heng-Yi Yuan\thanksref{e2,addr1}
        \and
        Meng-Hua Zhou\thanksref{e3,addr1}
        \and
        Kun Lu\thanksref{e4,addr1}
        \and
        Jing-Yi Wu\thanksref{e5,addr2}
        \and
        Kilar Zhang\thanksref{e6,addr1,addr3,addr4}
}

\thankstext{e1}{e-mail: clf@shu.edu.cn}
\thankstext{e2}{e-mail: natalie1225@shu.edu.cn}
\thankstext{e3}{e-mail: ywzmhatd@shu.edu.cn}
\thankstext{e4}{e-mail: 2858446139@shu.edu.cn}
\thankstext{e5}{e-mail: wujingyi222@mails.ucas.ac.cn; corresponding author}
\thankstext{e6}{e-mail: kilar@shu.edu.cn; corresponding author}

\institute{Department of Physics and Institute for Quantum Science and Technology, Shanghai University, Shanghai 200444, China\label{addr1}
          \and
          School of Astronomy and Space Science, University of Chinese Academy of Sciences (UCAS), Beijing 100049, China\label{addr2}
          \and
          Shanghai Key Lab for Astrophysics, Shanghai 200234, China\label{addr3}
          \and
          Shanghai Key Laboratory of High Temperature Superconductors, Shanghai 200444, China\label{addr4}
}

\date{}

\maketitle

\begin{abstract}
We extract the equation of state of hot quark matter from a holographic $2+1$ flavor QCD model, which could form the core of a stable compact star. By adding a thin hadron shell, a new type of hybrid star is constructed. With the temperature serving as a parameter, the EoS varies and we obtain stable stars with mass ranging from about 5 to 30 solar masses, and the maximum compactness around $0.2$. The I-Love-Q-C relations are further discussed, and compared with the neutron star cases. These compact stars are candidates for black hole mimickers, which could be observed by gravitational waves and distinguished by properties like nonzero tidal Love number and electromagnetic signals.
\end{abstract}

\section{Introduction}\label{sec:1}
 The equation of state (EoS) for compact stars, especially neutron stars (NS), has been a long puzzle. It involves extensively non-perturbative calculation for high density neutrons, beyond the current knowledge of nuclear physics. A hopeful way to find some hints is through astronomical observations. Traditional electromagnetic (EM) observation can fix pulsar masses to high accuracy, and recently developed gravitational wave (GW) interferometers \cite{LIGOScientific:2017vwq,LIGOScientific:2018hze,LIGOScientific:2020aai} can decipher binary masses and the combined tidal Love number (TLN) \cite{Hinderer:2007mb}. In addition, new detectors like NICER \cite{Miller:2021qha} can give information on the radius to higher precision. All of these could place constraints on EoS by narrowing the window for the corresponding mass-radius curve. They could rule out half of the more than one hundred known EoS, but still not enough to break the degeneracies of the other half.

Since multi-messenger observations \cite{LIGOScientific:2017ync,Dietrich:2020efo} are offering some but not sufficient information for compact star EoS, it is necessary to derive it theoretically. Currently, there are widely used NS EoS candidates, such as SLy \cite{Douchin:2001sv}, APR \cite{Akmal:1998cf}, satisfying the observation constrains, but with many parameters. 
A way out is to apply the holographic QCD \cite{Witten:1998zw,Polchinski:2000uf}, a generalization of the AdS/CFT duality \cite{Maldacena:1997re}, initiated from superstring theory. In \cite{Witten:1998zw,Sakai:2005yt,Sakai:2004cn} the Witten-Sakai-Sugimoto (WSS) model is introduced, basing on a D4/D8 brane setup. In \cite{Zhang:2019tqd,Li:2024ayw} the EoS are extracted using instanton gas, and in \cite{Kovensky:2021kzl} from homogeneous ansatz. However, the former cannot reproduce first order phase transition, and the latter cannot have chiral restoration. Very recently, \cite{Liu:2024efy} uses the Einstein-Maxwell-dilaton (EMD) system and Einstein-Maxwell-dilaton-scalar (EMD$\chi$) models to investigate the 2-flaovr QCD phase diagram, and the latter can lead to NS model meeting the observation constraints.

Inside the core of NS, the high pressure may cause the quark to be exposed, and many studies on quark stars (QS) have been performed. In parallel to the D4/D8 brane setup, there are studies on quark EoS based on D3/D7 model \cite{Hoyos:2016zke,Hoyos:2016cob,Annala:2017tqz,BitaghsirFadafan:2019ofb}, which leads to the conclusion that the quark matter cannot exist in neutron stars. Besides, the EoS itself is "softer" than the NS EoS, making the corresponding stars less compact than a normal NS,
which is against the instinct.  In \cite{Zhang:2022uin}, within the EMD system and Karch-Katz-Son-Stephanov (KKSS) framework, parameters from model in \cite{Cai:2022omk} are utilized to study the EoS for cold dense matter inside NS. It was also found that quark matter in NS is unstable. There are indeed results on hybrid stars with stable cold quark cores with special  setup \cite{BitaghsirFadafan:2020otb}, or predicted by observations and the sound speed \cite{Annala:2019puf,Han:2022rug,Fan:2023spm}, and also some studies on stable strange quark matter stars \cite{Kurkela:2009gj}.

 A way out is to consider another possibility. Rather than cold quark matter under high pressure, there could also be hot quark matter with lower pressure.
 In \cite{Aoki:2006we,Bazavov:2009zn}, studies on hot quark matter are performed, with the state equation and QCD phase transition given by lattice QCD method. In \cite{Zollner:2024iza}, warm EoS is obtained from the EMD model. In \cite{Cai:2022omk}, a holographic model for $2+1$ flavor QCD compatible with lattice QCD data is proposed (and in \cite{Zhao:2023gur} the 2-flavor case), with finite chemical potential and finite temperature. In this paper, we extract the EoS for quark-gluon plasma phase from this 2 + 1 flavor model, which could form a quark core of a compact star. {For the crust, it may cool down to hadron phase, where we could use a (hot) neutron-like EoS.} We then discuss the mass-radius and I-Love-Q-C relations \cite{Yagi:2013awa,Yagi:2013bca,Yagi2014WhyIE} of this two-layer model, following the method in \cite{Zhang:2020pfh}. It is possible to interpret some GW events as this kind of hot QS, which on the contrary will constrain the parameters in this model.

Since now we have two variables, the chemical potential $\mu$ and the temperature $T$, for simplicity, we choose five different constrains between the two, to illustrate the wide parameter space. We use fixed ratios of $\mu/T$, fixed $\mu$  and fixed $T$, around the critical end point (CEP). The maximum masses of the resultant stars lie between 23 to 30 $M_{\odot}$ (solar mass), with the compactness around $0.2$. The minimum masses are about 5 $M_{\odot}$.  These belong to a totally new kind of compact stars, very different from NS, mainly resulting from their lower central pressure and higher temperature.

The potential observation of this kind of hot QS is hopeful. Judging from their masses, they could be detected as black hole (BH) mimickers \cite{Lemos:2008cv,Cardoso:2019rvt} through GW observatories.
{BH mimickers are exotic compact objects that imitate many observable properties of BH such as extreme compactness, strong gravitational fields, and sometimes similar GW or electromagnetic signatures, but do not possess event horizons. They could be boson stars \cite{Liebling:2012fv}, gravastars \cite{Mazur:2001fv}, fuzzballs\cite{Mathur:2005zp} or QS, etc, since their existence cannot be easily excluded from observational data in favor of BH \cite{Abramowicz:2002vt}.}

Hot QS has a significant difference from BH: the low compactness will lead to high TLN and different wave forms, rather than zero TLN for BH. {Thus, it will be not difficult to distinguish hot QS if one could gain access to the TLN information, which GW interferometers like LIGO/Virgo could provide \cite{LIGOScientific:2017vwq}. In binary systems of compact stars, the GW waveform encodes the tidal deformability of the components. Instead of measuring the individual TLN of the two stars, a combined tidal deformability parameter defined as \\
$\tilde{\Lambda}=\frac{16}{13}\frac{(M_1+12M_2)M_1^4 \Lambda_1 + (M_2+12M_1)M_2^4 \Lambda_2}{(M_1+M_2)^5}$ is accessible, where $M_1$ and $M_2$ are the masses of the two stars, and $\Lambda_1$ and $\Lambda_2$ are their respective TLN \cite{Flanagan_2008,Hinderer_2010}. The tidal deformability of a hot QS–NS binary differs markedly from that of a BH–NS binary with the same NS companion. For example, for a binary consisting of a 20 $M_\odot$ BH and a 2 $M_\odot$ NS (with $\Lambda \approx 20$), the combined deformability is $\tilde{\Lambda} = 0.02$, which is likely undetectable. In contrast, replacing the BH with a 20 $M_\odot$ hot QS (with $\Lambda \approx 1000$ as an example) yields $\tilde{\Lambda}=1681$. Such a large difference in tidal response would make it possible to distinguish QS from BH.} In this sense, it is meaningful to recheck the TLN in GW events with one companion less than 30 $M_{\odot}$, such as GW190413\_052954,
GW190630\_185205, GW170814,
GW190828\_063405, 
GW200209\_085452 \cite{PhysRevX.9.031040,PhysRevX.13.041039} and pay special attention to future events within this range. Furthermore, the high temperate quark matter in such hot QS will have characteristic EM effects (though its significance depends a lot on the distance from the earth), making it also very different from possible dark stars which are supposed to have hardly any EM signal.



The paper is organized as follows: Section~\ref{sec:1} is the introduction; in Section~\ref{sec:2} we show how to extract EoS from a known holographic QCD model; Section~\ref{sec:3} illustrate the properties of the hot QS detectable by GW; we conclude in section~\ref{sec:4}.

\section{Holographic QCD Model}\label{sec:2}
The no-hair theorem is violated in asymptotic AdS spacetime \cite{Torii:2001pg}, and the emergence of scalar hair contributes additional terms to the first law of BH thermodynamics \cite{Lu:2014maa}. In\cite{Gubser:2008yx}, a five-dimensional gravity model with a scalar field was considered, the potential of the scalar field is designed to reproduce the equation of state of QCD. Based on this, \cite{DeWolfe:2010he,DeWolfe:2011ts} simulates the phase transitions and critical phenomena of QCD at finite temperature and density. To construct the equation of state for neutron stars, we require the explicit forms of energy and pressure. Therefore, we chose another holographic QCD model. Research in \cite{Li:2020spf} investigates the thermodynamics of AdS BH under Einstein-scalar gravity, defining thermodynamic quantities and mass through holographic renormalization. In \cite{Cai:2022omk} scenarios with nonzero chemical potential are introduced, compatible with the lattice model with zero chemical potential. Our calculations directly employ the model in \cite{Cai:2022omk}. Unlike previous studies focusing on cold quark matter, in this model we consider quark-gluon plasmas around the CEP, with the temperature reaching as high as 100 MeV ($1 \rm{MeV} \approx 1.16\times 10^{10} \rm{K}$).



We start from the action of the five dimensional EMD theory, 
\begin{equation}
    \begin{split}
        S=&\frac{1}{2\kappa_N ^2}\int d^5x\sqrt{-g} \\
        &\left[R-\frac{1}{2}\bigtriangledown_\mu\phi\bigtriangledown^\mu\phi-\frac{Z(\phi)}{4}F_{\mu\nu} F^{\mu\nu}-V(\phi)\right], \label{eq 1}
    \end{split}
\end{equation}
where $\kappa_N$ is the effective Newton constant, $\phi$ is the scalar field, $Z(\phi)$ and $V(\phi)$ are the coupling constants determined from the lattice QCD data at $\mu=0$.
\begin{align}
    V(\phi)&=-12{\rm cosh}[c_1 \phi]+(6c_1^2-\frac{3}{2})\phi^2+c_2 \phi^6\notag, \\
    Z(\phi)&=\frac{{\rm sech}[c_4 \phi^3]}{1+c_3}+\frac{c_3}{1+c_3}e^{-c_5\phi}.\label{eq z}
\end{align}
The parameters $\kappa_N$ and $c_i$  are fixed \cite{Cai:2022omk} with lattice data,  that $\kappa_N^2=2\pi(1.68)$, $c_1=0.7100, c_2=0.0037, c_3=1.935, c_4=0.085$, and $c_5=30$.

Comparing to the original form in \cite{DeWolfe:2010he}, the form of $Z(\phi)$ exhibits a spurious spike at low $\phi$. This is due to the presence of the last term in \ref{eq z}, which is numerically zero. In \cite{Jokela:2024xgz}, it is suggested that this term could be safely removed to simplify the analysis. Here we stay with this form of $Z(\phi)$ in \cite{Cai:2022omk}, since it not only is obtained by comparing with lattice data consistently, but also can explicitly provides the energy density and pressure. Besides, although this spike may effect the results at high densities and low temperatures (corresponding to low $\phi$), in this paper we are dealing with the opposite region. 

The metric of the hairy black hole construction is 
\begin{equation}
    ds^2=-f(r)e^{-\eta(r)}dt^2+\frac{dr^2}{f(r)}+r^2(dx^2+dy^2+dz^2).
\end{equation}

In this holographic model, both the energy density $\epsilon$ and the pressure $p$ can be represented by parameters on the AdS boundary,
\begin{align}
    \epsilon&=\frac{1}{2\kappa_N^2}(-3f_v+\phi_s \phi_v+\frac{1+48b}{48}\phi_s^4), \notag\\
    p&=\frac{1}{2\kappa_N^2}(-f_v+\phi_s \phi_v+\frac{3-48b-8c_1^4}{48}\phi_s^4),\label{eq 2}
\end{align}
with $b=-0.27341$, as a result of holographic renormalization. We can take the data on the horizon as an initial condition, solve the differential equation system with the NDSolve numerical method, then apply the numerical solution to fit the parameters.

With the above thermodynamic variables, the feasibility of constructing a QS can be considered. Unlike typical NS, it has the property of high temperature and low density.  Moreover, since the minimum pressure of the hot quark phase is nonzero,
 we need to add a layer of (also hot) hadron phase (or other possible components) externally, in order to construct a reasonable celestial body. 

In QCD the common units of energy density and pressure are both $[\rm{MeV}^4]$, when performing subsequent calculations, they need to be converted to astrophysical units: $r_{\odot}=G_N M_{\odot} / c^2, \epsilon_{\odot}=M_{\odot} /r_{\odot}^3$, and $p_{\odot}=c^2 \epsilon_{\odot}$, and we have
\begin{align}
    \rm{MeV}^4&=935340 \rm{kg}/\rm{m}^3=1.51453\times10^{-15}\epsilon_{\odot}\notag, \\\rm{MeV}^4&=8.40643\times10^{22}\rm{kg}/(\rm{s}^2 \rm{m})=1.51453\times10^{-15} \textit{p}_{\odot}.
\end{align}

\begin{figure*}[htb]
\includegraphics[width=\textwidth]{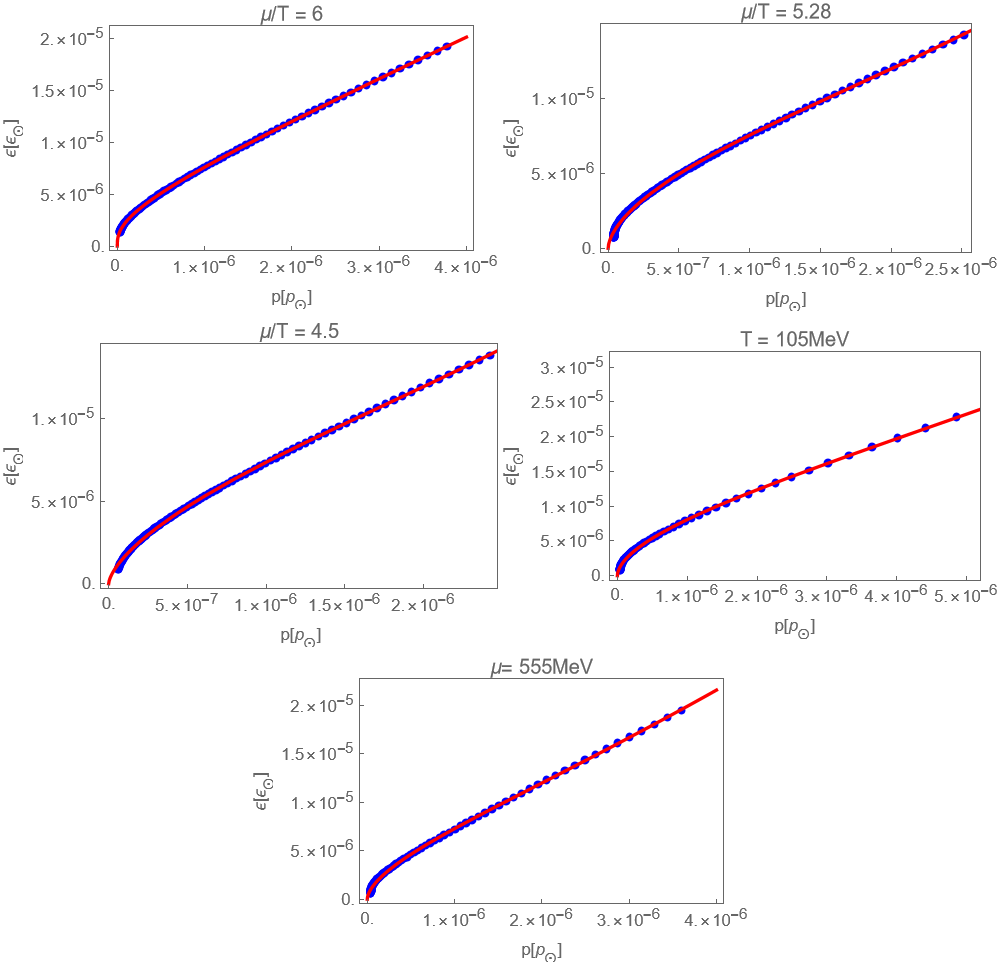}
\caption{The energy-pressure data and fitting curves under five different ${T,\mu}$ conditions. The blue dots represent the numerical data, and the red lines are the fitting curves. The data near the origin have small blanks, which indicate reaching the phase transition point.}
\label{fig:enter-label}
\end{figure*}

We can then extract the EoS: comparing to the cold quark models which usually use zero temperature, now
$T$ serves as an extra parameter, together with the chemical potential $\mu$. Since the specific relationship for $\mu$ and T as functions of the position in the star is complicated and unknown, we choose to fix the ratio between $T$ and $\mu$ to grab some general information of the parameter space, 
rather than discussing their detailed relation with the radius. Since the CEP obtained in \cite{Cai:2022omk} is at $T=105\rm{MeV}$ and $\mu=555\rm{MeV}$, we pick five typical data, namely $\mu/T=4.5$, $\mu/T=5.28$, $\mu/T=6$, $T=105\rm{MeV}$ and $\mu=555\rm{MeV}$, to extract the corresponding EoS. {For the data where T is fixed, we chose $\mu\in [555\rm{MeV}, 1440\rm{MeV}]$, for the data where $\mu$ is fixed, we chose $T\in [105\rm{MeV}, 180\rm{MeV}]$}. Their energy density-pressure data graphs and corresponding fitting curves are drawn respectively in Fig.\ref{fig:enter-label}.  

\begin{figure*}[htb]
    \centering  
    \includegraphics[width=0.7\linewidth]{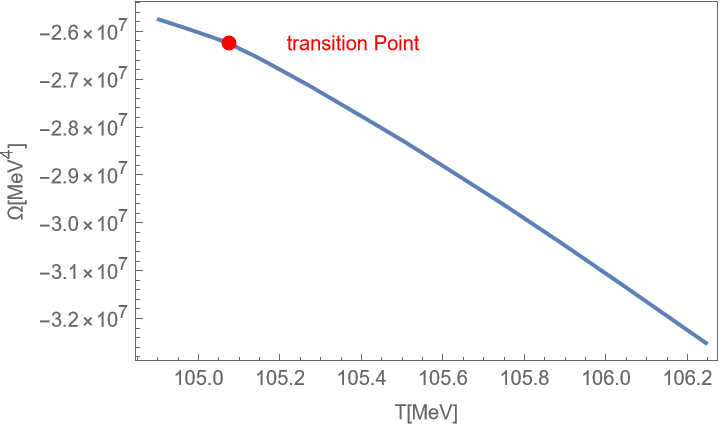} 
    \caption{Curve of free energy and temperature at $\mu=555MeV$, where the red point indicates not only a phase transition point, but also the QCD critical endpoint.}
    \label{555MeV}
\end{figure*}

\begin{figure*}[htb]
    \centering  
    \includegraphics[width=0.7\linewidth]{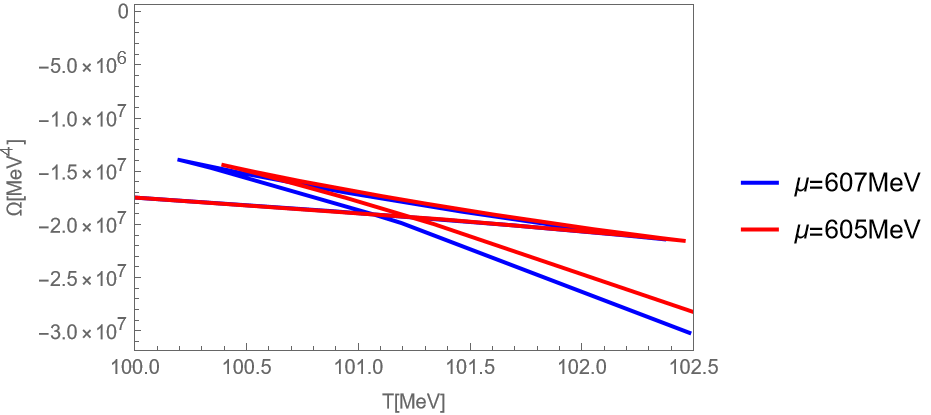} 
    \caption{Curves of free energy and temperature at $\mu/T=6$. There is a point of intersection in the curves, which corresponds to the phase transition point. This part of the multi-valued function means that the first order phase transition occurs. The red curve corresponds to the case of $\mu=605MeV$, and above this, the points have entered the phase transition region, while the blue curve $\mu=607MeV$ has not entered.}
    \label{phase diagram at 6}
\end{figure*}

There are some nuances we cannot ignore. See Fig.\ref{fig:enter-label}, for the quark phase, the pressure cannot drop to 0, but around $10^{-7} p_{\odot}$ (or about $10^{7} [\rm{MeV}^4]$), where the phase transition ({quark-gluon plasma phase transition to hadrons}) lies.
For the case of $T=105\rm{MeV}$, $\mu=555\rm{MeV}$, $\mu/T=5.28$ and $\mu/T=6$, they have obvious phase transition points, and the first three cases pass through CEP in Fig.\ref{555MeV}, so the CEP is directly used as the lower limit of the data ($p=3.97 \times 10^{-8} p_{\odot}=2.62127\times10^7 \rm{MeV}^4$). For $\mu/T=6$, it can be seen from Fig.\ref{phase diagram at 6} that at 605MeV, the data point is on the left of the phase transition point, and the data point of 607MeV is on the right, so that the lower limit of the data can be read off ($p=3.03394 \times 10^{-8} p_{\odot}=2.00322\times10^7 \rm{MeV}^4$). Although $\mu/T=4.5$ does not pass through the first-order phase transition line, we find a distinct high order phase transition point in the pressure-energy data, so we take the data before that point ($p=6.07854 \times 10^{-8} p_{\odot}=4.01348\times10^7 \rm{MeV}^4$).


Their fitting results ($p_1, \epsilon_1$ corresponds to $\mu/T=4.5$, $p_2, \epsilon_2$ corresponds to $\mu/T=5.28$, $p_3, \epsilon_3$ corresponds to $\mu/T=6$, $p_4, \epsilon_4$ corresponds to $T=105\rm{MeV}$, $p_5, \epsilon_5$ corresponds to $\mu=555\rm{MeV}$ ) are listed, respectively:
\begin{equation}
    \begin{split}
        \epsilon_1&=5.92293\times 10^{-2} \,p_1^{0.651263} + 1.43266\times 10^{13} \,p_1^{3.42233}, \\
        p_1&\in [6.07854\times 10^{-8}, \;2.42143\times 10^{-6}], \\ 
        \epsilon_1&\in[8.65502\times 10^{-7}, \;1.3846\times 10^{-5}], 
    \end{split}
    \label{eos1} 
\end{equation}

\begin{equation}
    \begin{split}
        \epsilon_2&=2.7042\times 10^{-2} \,p_2^{0.593438}+1.80685\times10^{10} \,p_2^{2.87626},\\ 
        p_2&\in [3.97089\times 10^{-8}, \;2.5162\times 10^{-6}], \\ 
        \epsilon_2&\in[7.09555\times 10^{-7}, \;1.42\times 10^{-5}],
    \end{split}
    \label{eos2}
\end{equation}

\begin{equation}
    \begin{split}
        \epsilon_3&=8.33022\times 10^{-4} \,p_3^{0.370279} +8.72125 \,p_3^{1.08721},\\ 
        p_3&\in [3.03394\times 10^{-8}, \;9.51759\times 10^{-6}],\\ 
        \epsilon_3&\in[1.36385\times 10^{-6}, \;1.9245\times 10^{-5}], 
    \end{split}
    \label{eos3}
\end{equation}

\begin{equation}
    \begin{split}
        \epsilon_4&=1.7254\times 10^{-2} \,p_4^{0.557614} +3.33051\times 10^{3} \,p_4^{1.68025},\\ 
        p_4&\in [3.97113\times10^{-8}, \;7.12588\times10^{-6}], \\ 
        \epsilon_4&\in[7.98295\times10^{-7}, \;3.05948\times 10^{-5}], 
    \end{split}
    \label{eos4}
\end{equation}

\begin{equation}
    \begin{split}
        \epsilon_5&=2.35768\times 10^{-2} \,p_5^{0.590736} +2.30251\times 10^{4}\,p_5^{1.77162}, \\ 
        p_5&\in [3.97265\times10^{-8}, \;3.60026\times10^{-6}], \\ 
        \epsilon_5&\in[6.111821\times10^{-7},\; 1.945\times 10^{-5}]. 
    \end{split}
    \label{eq:eos5}
\end{equation}
For the fitting range of $p$ and $\epsilon$ (in astrophysical units), the lower bound is set on the phase transition point, while the upper bound is chosen to ensure star stability.

Again we must emphasize that, none of the above 5 cases shows the real EoS for the QS, but they constrain the range where the real EoS could lie in.

\section{Hot Quark Star}\label{sec:3}
Now we have inferred some models of compact stars by holographic QCD, but an important question remaining is how to relate the theory to observations. In \cite{PhysRevD.88.023009}, the universal relations consisting of the momentum of inertial (I), the tidal deformability (Love) and the quadrupole moment (Q) were discovered. The I-Love-Q universal relations are meaningful for astrophysics, GW astronomy and fundamental physics.

\begin{figure*}
    \centering
    \includegraphics[width=\linewidth]{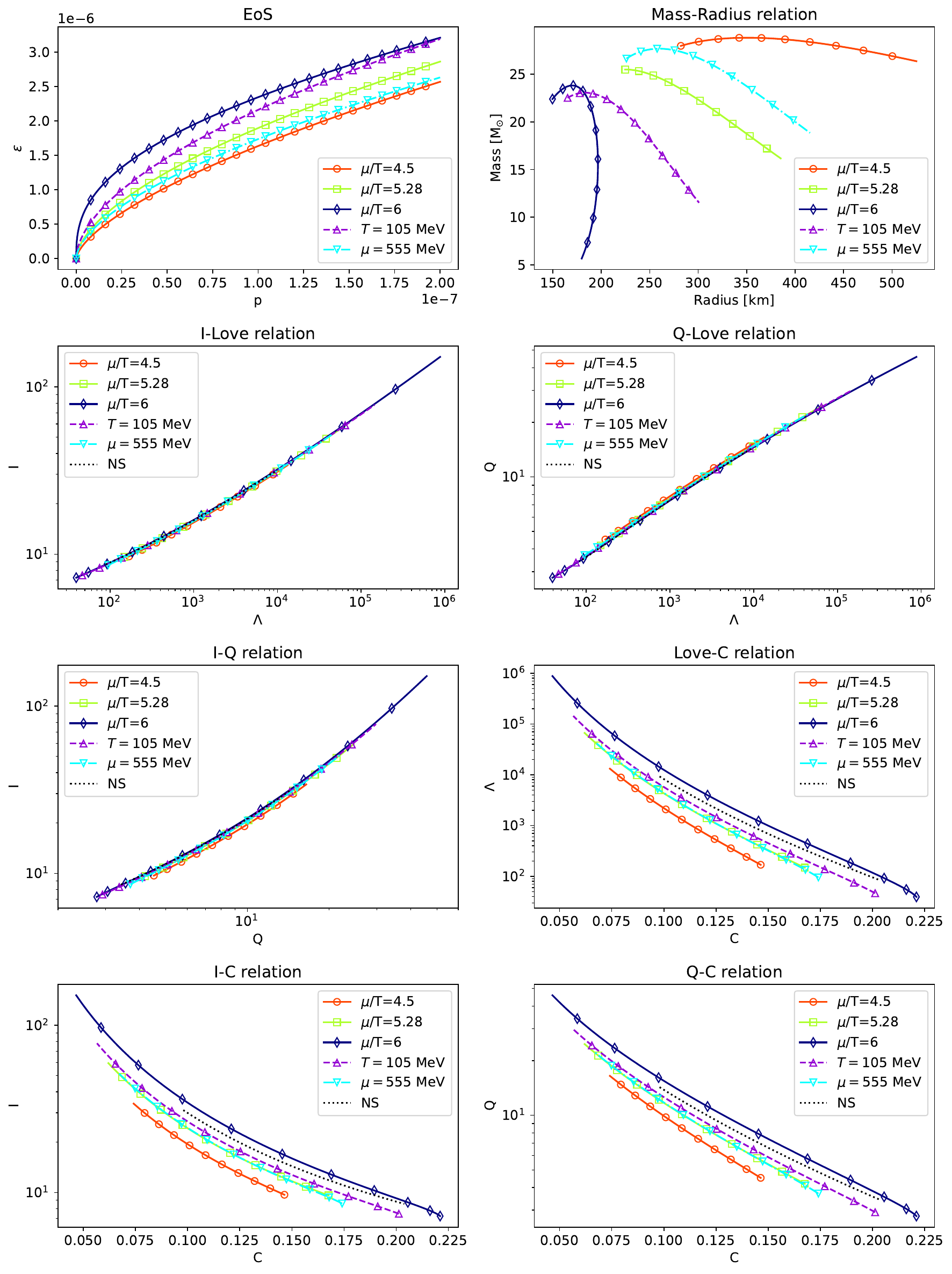}
    \caption{The EoS, M-R relation and I-Love-Q-C relations for "simple" quark star with different models in eq.(\ref{eos1}-\ref{eq:eos5}). For reference, the black dotted line representing a polytropic NS $\epsilon=0.09p^{0.5}$ is plotted in the panels of I-Love-Q-C relations.}
    \label{iloveqc}
\end{figure*}

Although the minimum quark-gluon plasma phase is nonzero, the approach to zero pressure is essential to form a star, because the zero pressure represents the star surface. 
The most naive way to find the EoS to zero pressure is to simply use the fitting EoS, as extrapolations of the original data, which are valid only larger than the phase transition values. This means a smooth phase transition from the quark core to the outer hadron layer, not realistic enough, but can serve as a reference.
In Fig.\ref{iloveqc}, we present the EoS and the I-Love-Q relations for this kind of "simple" QS models (this terminology are used afterward). Additionally, the M-R relation and the relations related to the compactness $C$ are shown, applying the
calculation procedures in \cite{PhysRevD.88.023009}. We obtain that, the maximum masses are about 23 to 30 $M_{\odot}$, and the compactness is about $0.1$.

{More rigorously, since there exists a phase transition from quark-gluon plasma core to the hadronic crust as the star cools, we should connect the holographic EoS derived above with a hot (the temperature is expected to drop toward the surface) neutron-like model to calculate the mass-radius relation and the I-Love-Q-C relations. For the sake of simplicity in illustrating the idea, we adopt a single-component polytropic EoS to model the outer layer}
\begin{equation}
    \qquad \qquad \qquad \qquad \epsilon_{ns} = \kappa_{ns} \,p^{\gamma_{ns}}.
\end{equation}
In this paper we fix $\gamma_{ns}=0.5$. It is natural that the energy density discontinues at the transition point, and we set
\begin{equation}
    \qquad \qquad \qquad \qquad \epsilon_{ns} = n \,\epsilon_{qs},
    \label{phasetran}
\end{equation}
where $n$ can be achieved by adjusting the parameter $\kappa_{ns}$. 
For the actual connection, one need the equivalence of the pressure, chemical potential and temperature at the same time. However, since the outer EoS is already an assumption that the detailed reliance of the $T$ and $\mu$ to $p$ is undetermined and adjustable, it is adequate to consider the phenomenological discontinuity of the $\epsilon$-$p$ curve for the current study.

\begin{figure*}
    \centering
    \includegraphics[width=\linewidth]{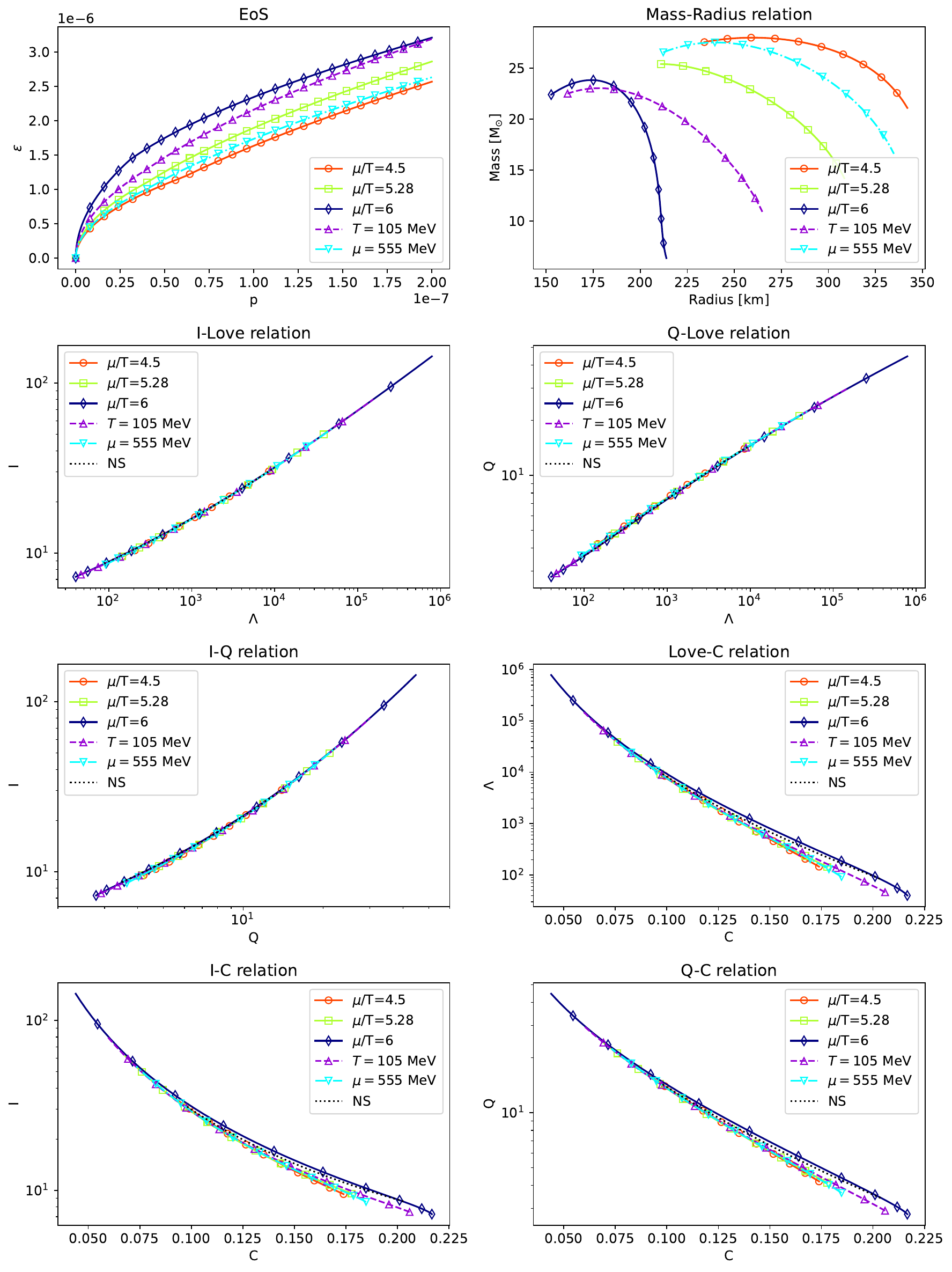}
    \caption{The EoS, M-R relation and I-Love-Q-C relations of the hybrid stars consist of quark cores and neutron outer layers with $n=1$ in eq.(\ref{phasetran}). For reference, the black dotted line representing a polytropic NS $\epsilon=0.09p^{0.5}$ is plotted in the panels of I-Love-Q-C relations.}
    \label{iloveqc1}
\end{figure*}

\begin{figure*}
    \centering
    \includegraphics[width=\linewidth]{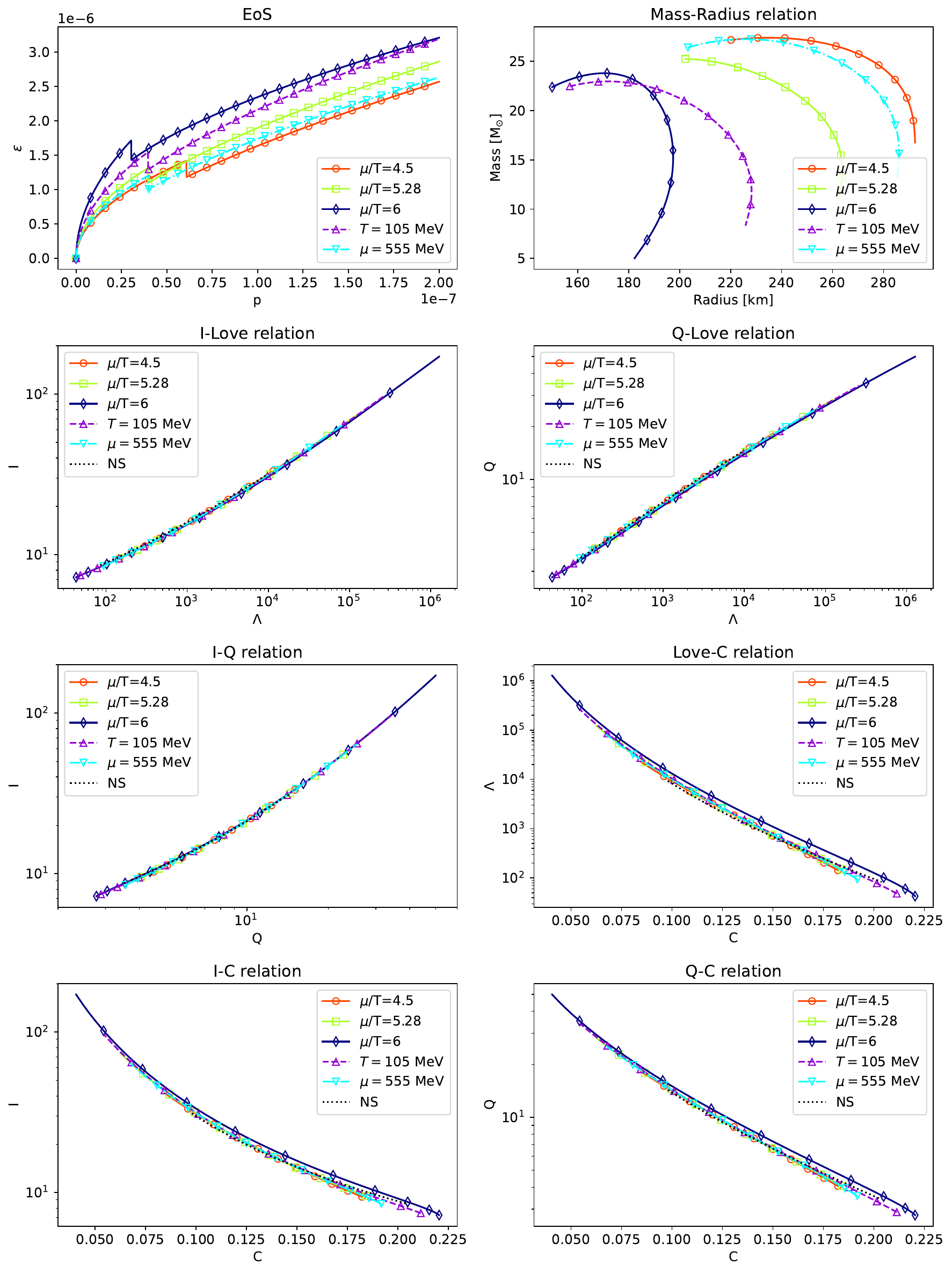}
    \caption{The EoS, M-R relation and I-Love-Q-C relations of the hybrid stars consist of quark cores and neutron outer layers with $n=1.2$ in eq.(\ref{phasetran}). For reference, the black dotted line representing a polytropic NS $\epsilon=0.09p^{0.5}$ is plotted in the panels of I-Love-Q-C relations.}
    \label{iloveqc12}
\end{figure*}

\begin{figure*}
    \centering
    \includegraphics[width=\linewidth]{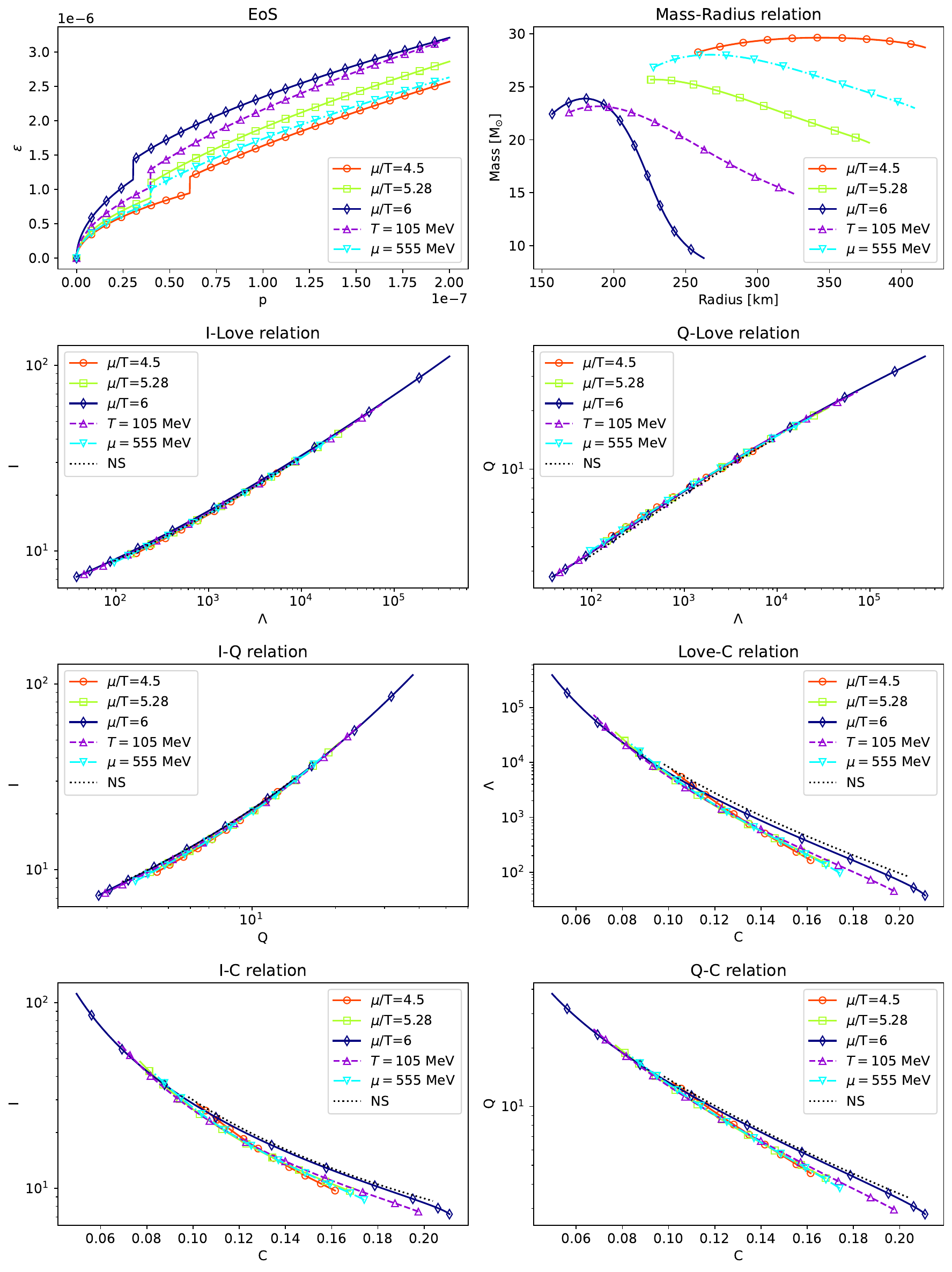}
    \caption{The EoS, M-R relation and I-Love-Q-C relations of the hybrid stars consist of quark cores and neutron outer layers with $n=0.8$ in eq.(\ref{phasetran}). For reference, the black dotted line representing a polytropic NS $\epsilon=0.09p^{0.5}$ is plotted in the panels of I-Love-Q-C relations.}
    \label{iloveqc08}
\end{figure*}

\begin{figure*}
    \centering
    \includegraphics[width=\linewidth]{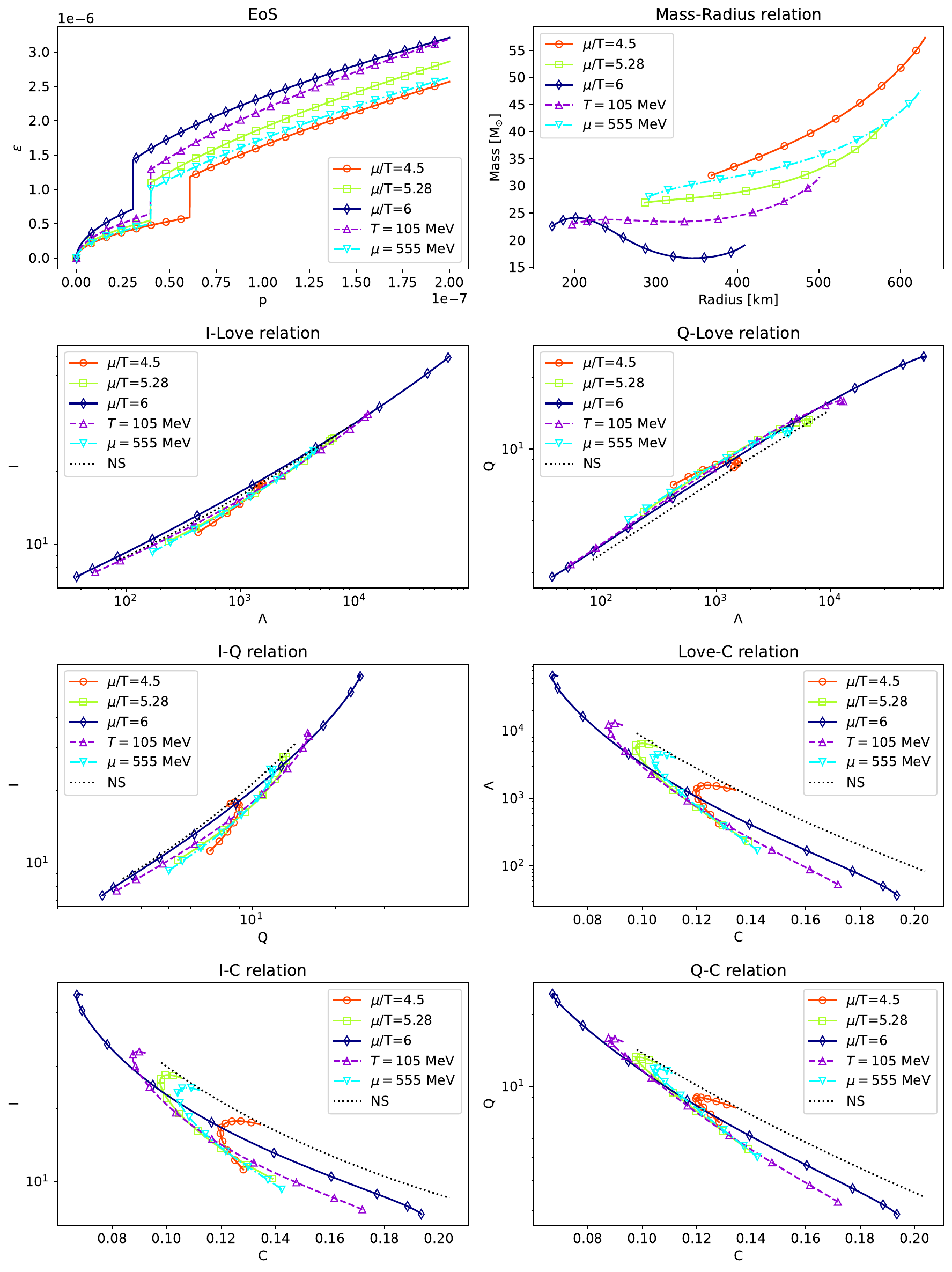}
    \caption{The EoS, M-R relation and I-Love-Q-C relations of the hybrid stars consist of quark cores and neutron outer layers with $n=0.5$ in eq.(\ref{phasetran}). For reference, the black dotted line representing a polytropic NS $\epsilon=0.09p^{0.5}$ is plotted in the panels of I-Love-Q-C relations.}
    \label{iloveqc05}
\end{figure*}

In Fig.\ref{iloveqc1}-\ref{iloveqc05}, we show the corresponding curves for hybrid models with $n=1$, $1.2$, $0.8$ and $0.5$, respectively, following the two-layer hybrid star calculation method in \cite{Zhang:2020pfh}. For reference,  the I-Love-Q-C relations of an NS model with $\epsilon=0.09\,p^{0.5}$ are indicated by the black dotted lines\footnote{As a result of the scaling symmetry \cite{Wu:2023aaz,Zhang:2023hxd,Maselli:2017vfi}, $\epsilon=\kappa p^{\gamma}$ with different $\kappa$ all have the same curves.}. 

{Notice that all the curves (the M-R relations etc.) in Fig.\ref{iloveqc1}-\ref{iloveqc05} (except the left-top panels for EoS) are illustrated with the pressure ranges given in \eqref{eos1} to \eqref{eq:eos5} , which means the lowest star central pressures lie on the phase transition points from quark-gluon to hadrons. Below those chosen lowest pressure values, the stars will be no longer QS  but pure hadronic, without any quark-gluon state, and in our case the corresponding EoS for pure hadrons are just single polytropic, whose behaviors are well-known. All the stars represented in the curves have quark cores, while stars made by pure hadrons are not drawn, so no apparent phase transitions are shown, in contrast to those in the left-top panel EoS curves.} 

In addition to the discontinuous EoS with $n=0.5$ in Fig.\ref{iloveqc05}, it already demonstrates that the I-Love-Q trio are broken even in the continuous model in Fig.\ref{iloveqc}. Actually, the results from \cite{Wu:2023aaz} indicate that the relations for extreme continuous EoS are also broken, in addition to the cases  verified in many previous papers \cite{Yagi2014WhyIE,Haskell:2013vha,Silva2017ILoveQTT,PhysRevD.103.023022,Li:2023owg,Atta:2024ckt,Kumar:2023ojk,Pani:2015tga}. Regarding the I-C, Love-C and Q-C relations, the universality is clearly lost. However, as pointed out in \cite{PhysRevD.88.023009}, the loss of universality in compactness relations for realistic NS occurs. Also in \cite{Hoyos:2021uff}, the authors analytically obtain the quasi-universal relations of a holographic quark eos, which lends support to the rationality of our findings. 

By comparing the panels for M-R relations in these figures, we observe that the hybrid stars become unstable for large phase transitions, as shown in Fig.\ref{iloveqc05}. Interestingly, the maximum M-R values for each model appear to match those of the "simple" quark case, though deviations become more pronounced when $\mu/T=4.5$. To investigate the cause, we calculate the maximum mass and radius at the phase transition point for each model (the quark core) in Table.\ref{coremr}. The M-R pairs for the cores are listed in the second column, while the total star M-R pairs for "simple" QS are provided in the last column. It is evident that all models exhibit heavy cores (with the exception of $\mu/T=4.5$ case), and the M-R relations are significantly influenced by the core properties.  

The panels of I-Love-Q-C relations show 
that the universal relations are observed when applying connected EoS, with the exception of $\mu/T=4.5$. In \cite{PhysRevD.88.023009}, a possible explanation for the existence of I-Love-Q relations is suggested, namely that the EoS of realistic NS models approach each other in the outer layers. Since we connect the same polytropic models to various quark models, it is plausible that the deviations in the I-Love-Q-C relations for "simple" QS are reduced, even though these stars possess relatively light shells.    

In conclusion, the mass distribution of both "simple" QS and hybrid stars, all consisting of quark cores and hadron outer layers, allows these objects to mimic stellar-mass BH. The obedience of I-Love-Q-C relations for hybrid models presented in this paper extends the validity of these universal relations. On the front of observation, the I-Love-Q-C relations offer a potential means to distinguish between BH and QS because of the zero TLN for BH in GR. Furthermore, the broken relations about compactness help identify the holographic quark models. Since the I-Love-Q-C relations are valid for most of the hybrid models studied in this paper, these stars could potentially be distinguished from the "simple" quark cases. Additionally, the observation on the hybrid stars could place constraints on the phase transition from quark matter to hadronic matter when applying such models.  

\begin{table}
    \renewcommand\arraystretch{1.5}
    \centering
    \begin{tabular}{|c|c|c|}
        \hline
        $\quad$ & core & total \\
        \hline
        $\mu/T=4.5$ & $\left \{ 19, 182 \right \}$ & $\left \{ 29, 351 \right \}$\\
        $\mu/T=5.28$ & $\left \{ 23, 168 \right \}$ & $\left \{ 25, 225 \right \}$ \\
        $\mu/T=6$ & $\left \{ 22, 156 \right \}$ & $\left \{ 24, 171 \right \}$ \\
        $T=105\ \rm{MeV}$ & $\left \{ 21, 150 \right \}$ & $\left \{ 23, 183 \right \}$ \\
        $\mu=555\ \rm{MeV}$ & $\left \{ 23, 185 \right \}$ & $\left \{ 28, 261 \right \}$ \\
        \hline
    \end{tabular}
    \caption{$\left \{ M \  [M_{\odot}], R \  [\rm{km}] \right\}$ of quark core compared with the ones of total stars.  We only list the data of the maximum mass for each model in Fig.\ref{iloveqc}.}
    \label{coremr}
\end{table}

\section{Conclusions}\label{sec:4}
We extract the EoS of quark-gluon plasma at finite temperature using the holographic $2+1$ flavor QCD model and applied it to the construction of compact stars. Since the energy density and pressure are related to chemical potential and temperature, we selected five different cases of ${\mu, T}$ for analysis. Considering that the quark-gluon plasma in this model cannot reach a state of zero pressure, we added an outer layer of hadrons, connected by phase transition. We found that this approach can result in stable compact stars if the phase transition range is in a proper limit.

Applying the five connected EoS, we calculate the M-R relation and the I-Love-Q-C relations. Our analysis on the M-R relation, along with Table.(\ref{coremr}) which lists the maximum mass-radius values for both the core and the whole of each model, reveal that most models have heavy cores, leading to maximum masses that are similar to those of "simple" QS. The I-Love-Q-C panels show violations for "simple" QS especially with $\mu/T=4.5$. However, the presence of thin hadron shells sharing the same EoS mitigates these violations, as the outer layers of realistic NS EoS exhibit similarities.

The results have significant implications for future observations. First, the maximum masses of the hot QS are around 23 to 30 $\rm{M_\odot}$, which suggest that they could be observed in GW events as BH mimickers. Since BH exhibit zero TLN in GR, one could distinguish hot QS from BH through the inference on TLN. Moreover, the deviation from universality in the I-Love-Q-C relations for "simple" QS provides a means to infer the underlying quark model. Comparing with the I-Love-Q-C relations for the "simple" QS, the universal relations for the hybrid models remain valid, enabling the potential discrimination between them, verifying the hypothesis in \cite{Wu:2023aaz} that the outer layer contributes most to the universal relations. The universal relations could also be employed to constrain the parameters in the phase transition. 

The characterizations of the hybrid stars are related to GW observations indirectly, but may still be constrained. For the phase transition models, the crucial parameters are $T-\mu$ and $\epsilon-p$ which also means $\kappa-\gamma$. The GW events may help constrain the parameters.

Our results have another interpretation. For the quark core, the pressure does not necessarily need to drop to the phase transition point, since it is possible for the outer layer to be other components like dark matter. But the calculation and analysis are the same, as long as the outer EoS is given. The EM signatures of the hot QS differ from those of the possible (pure or hybrid) dark stars, offering another avenue for distinguishing between these two types of stars through future multi-messenger observations.

For the next step, one quick direction is to consider the 2-flavor QCD case discussed in \cite{Zhao:2023gur}, with some progress given in \cite{Chen:2025ppr}. On the other hand, we will pay more attention on the detailed structure of the hot hadron outer layer. Moreover, it will be interesting to study the constructions of these hybrid stars when changing the outer layer EoS. 

The formation mechanism of hot QS and their lifetime are also interesting. 
One possibility is that it originates from the high-temperature early universe, while other possibilities include the collapse of a massive star, along with the binary mergers. Since a supernova explosion or binary merger can reach a temperature of 50 MeV \cite{Sekiguchi:2011mc,Drago:2015dea}, there is a good chance to hold a hot QS if this magnitude could double, achieving a quark state. This state could resist gravity through thermal motion and degeneracy pressure, preventing the formation of a BH. In fact, its relatively large radius results in a compactness of approximately $0.1$, which is lower than the typical NS compactness of $0.2$ and much smaller than the BH compactness of $0.5$. 

Considering the high temperature of the quark core, the radiation seems to be quite strong. However, 
the model we are discussing is not a bare quark star but rather one where the quark state forms the core, while the shell consist of hadronic matter that has undergone a phase transition. As we move toward the star’s surface, the temperature gradually decreases, leading to a reduction in the star’s overall radiation. On the other hand, at the boundary between the core and the outer layers (i.e., the phase transition region), radiation can be reflected back, maintaining a high-temperature core and a low-temperature outer shell. Radiation might indeed cause such a QS to have a relatively short lifespan, which could explain why no observational evidence has been found so far (or they might have been mistaken for BH, and due to their distance, tidal deformations or electromagnetic effects have not been detected). As the temperature falls, there will be a second collapse, which can be distinguished from a standard one-time explosion supernova. We can further discuss the possible existence time of this type of QS. The details are beyond the scope of the current paper, and will be left for future work.

$\\$
\noindent {\it Acknowledgements.} 
The authors thank Song He, Chian-Shu Chen, Alessandro Parisi, Chen Zhang and Zhoujian Cao for very helpful discussions.  K.Z. (Hong Zhang) is supported by a classified fund from Shanghai city.

\bibliographystyle{JHEP}
\bibliography{hQCD}

\end{document}